\documentclass{aastex62}
\usepackage{amsmath, amssymb, amsfonts}
\usepackage{natbib}

\newcommand{\beq}{\begin{equation}}
\newcommand{\eeq}{\end{equation}}

\graphicspath{{./}{figures/}}

\received{September 22, 2019}
\revised{April 9, 2020}
\accepted{April 10, 2020}
\shorttitle{Mass Function of NGC 1711}
\shortauthors{Madaan et al.}
\begin{document}

\title{Using the Modified Lognormal Power-Law Distribution to Model the Mass Function of NGC 1711}

\correspondingauthor{Shantanu Basu}
\email{basu@uwo.ca}

\author{Deepakshi Madaan}
\affiliation{Department of Applied Mathematics, The University of Western Ontario, London, ON N6A 5B7, Canada.}
\affiliation{Department of Physics and Astronomy, The University of Western Ontario, London, ON N6A 3K7, Canada.}

\author[0000-0003-3212-2216]{Sophia Lianou}
\affiliation{IAASARS, National Observatory of Athens, Penteli 15236, Greece.}
\affiliation{Department of Physics and Astronomy, The University of Western Ontario, London, ON N6A 3K7, Canada.}

\author[0000-0003-0855-350X]{Shantanu Basu}
\affiliation{Department of Physics and Astronomy, The University of Western Ontario, London, ON N6A 3K7, Canada.}
\affiliation{Department of Applied Mathematics, The University of Western Ontario, London, ON N6A 5B7, Canada.}



\begin{abstract}
A determination of the mass function (MF) of stellar clusters can be quite dependent on the range of measured masses, the fitting technique, and the analytic function that is being fit to the data. Here, we use Hubble Space Telescope/WFPC2 data of NGC 1711, a stellar cluster in the Large Magellanic Cloud, as a test case to explore a range of possible determinations of the MF from a single dataset. We employ the analytic modified lognormal power-law (MLP) distribution, a hybrid function that has a peaked lognormal-like body and a power-law tail at intermediate and high masses. A fit with the MLP has the advantage that the resulting best-fit function can be either a hybrid function, a pure lognormal, or a pure power law, in different limits of the function. The completeness limit for the observations means that the data contains masses above $\sim 0.90\,M_{\odot}$. In this case, the MLP fits yield essentially a pure power-law MF. We demonstrate that the nonlinear regression/least-squares approach is not justified since the underlying assumptions are not satisfied. By using maximum-likelihood estimation, which is independent of binning, we find a best-fit functional form $dN/d\ln m \propto m^{-\alpha}$, where $\alpha = 1.72 \pm 0.05$ or $1.75 \pm 0.05$ for two different theoretical isochrone models, respectively. Furthermore, we explore the possibility of systematic errors in the determination of the power-law index due to the depth of the observations. When we combine the observational data with artificially generated data from the lognormal Chabrier initial MF for masses below $0.90\, M_{\odot}$, the best-fit MLP is a hybrid function but with a steeper asymptotic slope i.e., $\alpha = 2.04 \pm 0.07$. This illustrates the systematic uncertainties in commonly used MF parameters that can depend on the range of data that is fitted.
\end{abstract}

\keywords{stars: luminosity function, mass function -- Magellanic Clouds -- methods: data analysis -- methods: statistical}


\section{Introduction} \label{sec:intro}

For any star with a given chemical composition, its initial mass determines its subsequent evolution \citep[e.g.,][]{Chiosi1992}. Once the mass of the star is known, various stellar properties can be derived, such as the luminosity, radius, and radiation spectrum. Also, various integrated properties of any group of stars, i.e., a star cluster or a galaxy, depends on how stellar masses are distributed into different mass intervals \citep{Scalo1986}. Hence, it is necessary to study the distribution of stellar masses at birth, known as the initial mass function (IMF), in order to understand the evolution of galaxies and their constituent stellar populations. The functional form of the IMF is significant since it is used as an input in stellar population synthesis analyses \citep[e.g.,][]{Buzzoni1989,Maraston1998,Bruzual2003,Kotulla2009}. The IMF enters into the equations to study the chemical evolution of galaxies \citep{Tinsley1980}, and to determine their star formation rate \citep{Kennicutt1998}. Predictions of luminosity functions (LFs) of white dwarfs \citep{Dantona1978} and the rate of formation of planetary nebulae \citep{Papp1983} also depend on the form of the IMF. Altogether, multiple astrophysical studies depend on the assumption of the functional form of the IMF, and many of them are very sensitive to its high-mass power-law index, since high-mass stars synthesize heavy elements and dominate radiative and mechanical energy feedback to their host galaxies. It is important then to study causes of variation in the measured IMFs, especially in their power-law tail. It is also very practical to employ a simple analytic and integrable form of the IMF so that one can explore its parameter space in an efficient manner.

Star formation occurs through a highly complex transformation of interstellar molecular clouds, and is controlled by various physical mechanisms such as self-gravity, turbulence, and magnetic fields \citep{Mous1999,Bonnell2007,Klessen2011,Offner2014}. Due to the stochastic nature of star formation, the mass of a star can be considered to be a continuous random variable and hence the fraction of stars in each mass interval formed at birth, i.e., the IMF, can be modeled as a probability density function (PDF). 
\citet{Salpeter1955} was the first to provide a functional fit to the IMF, i.e., $dN/d\ln m \propto m^{-\alpha}$, with index $\alpha=1.35$, by studying the LF of the main-sequence stars of masses $m \gtrsim 1\,M_{\odot}$ in the solar neighborhood. Subsequently, on finding that the observed stellar mass distribution flattens for low-mass stars, \citet{Miller1979} suggested a lognormal form below $1\,M_{\odot}$. \citet{Zinnecker1984} gave a theoretical explanation for a lognormal IMF by invoking the Central Limit Theorem (CLT). According to the CLT, the sum of a large number of independent and identically distributed random variables will follow a Gaussian distribution \citep{Aitchison1957}. Since the process of star formation is controlled by many physical processes, the formation of stellar masses can be considered to be a product of a large number of independent and identically distributed random variables. Thus, by the CLT, the log of the product of the random variables will follow a Gaussian distribution, implying that the stellar mass follows a lognormal distribution. \citet{Chabrier2005} also compiled observational data and found a lognormal fit for the substellar and low mass stellar regime, while adopting a power-law approximation with index $\alpha = 1.35$ for the intermediate and high mass stellar regime $m > 1\,M_{\odot}$. In a comprehensive study, \citet{Scalo1986} found a power-law index $\alpha =1.7$ for the intermediate and high mass regime $2\,M_{\odot} < m  < 10\,M_{\odot}$ for the IMF and also quoted the same index for a cluster-averaged IMF. Furthermore, \citet{Kroupa2001,Kroupa2002} used a multisegment power-law profile for the ranges 0.01 $M_{\odot}$ to 0.08 $M_{\odot}$, 0.08 $M_{\odot}$ to 0.50 $M_{\odot}$, and above 0.50 $M_{\odot}$. The best-fit index for $m > 0.5\,M_{\odot}$ was 1.3, but \citet{Kroupa2002} suggested that it could be closer to 1.7 when biases for unresolved binary systems were taken into account. All of these studies are a result of compiling information from a wide range of star-forming regions, and there is no guarantee that the IMF is universal and that every star-forming region follows the same parametric distribution i.e., a PDF with a fixed set of parameters. \citep[][see also \citet{Dib2017}]{Dib2014} found that there is a statistically significant variation of parameters that characterize the shape of the IMF, among a set of Galactic young stellar clusters. \citet{Dib2018} also showed that cluster-to-cluster variations of the intrinsic IMF could lead to a composite mass function that resembles the Galactic field IMF determined by \citet{Kroupa2002}.
 
\citet{Basu2004} introduced a hybrid three-parameter PDF, the modified lognormal power-law (MLP) distribution function, to model the entire stellar mass regime as a single function. Many other parameterized approximations need some sort of joining condition to connect different segments. This adds to the number of parameters involved. The functional form of \citet{Chabrier2005} has four parameters, including the joining condition. The multisegment power-law profile of \citet{Kroupa2002} is a five-parameter PDF, including the joining conditions. The MLP on the other hand does not require a joining condition and is a function of only three parameters, one more than the lognormal. As Chabrier suggested, the IMF can be fitted by a lognormal distribution with a characteristic peak and turnover for the low-mass stellar regime, and by a power law for the intermediate and high-mass regime. The MLP can model the entire mass regime with just three parameters, showing both lognormal-like and power-law-like behavior, and also reducing to a pure power law or lognormal if the data are consistent with those choices.

The MLP emerges from a generative model that is based on accretion termination, as described briefly in \S\ \ref{MLPsection} and in more detail in \citet{Basu2015}. Accretion termination and turbulent fragmentation are two opposing paradigms for the origin of the IMF, although other scenarios are certainly also possible. There is a rich set of literature on each topic. Accretion termination has been proposed to be caused by protostellar outflows \citep{Shu1987,Adams1996} and also by the ejection of protostellar embryos from disks \citep{Bate2002,Stamatellos2009,Basu2012}. A competing paradigm is that of turbulent fragmentation, in which turbulent fluctuations can cause a wide range of fragment masses to become unstable to gravitational collapse \citep{Padoan2002,Hennebelle2008,Hennebelle2009}. The accretion termination picture is a bottom-up creation process that starts from very low-mass seeds that grow by accretion, whereas the turbulent fragmentation is a top-down process in which turbulence, magnetic fields, and gravity set the collapsing masses out of which a significant portion finds itself in a star or brown dwarf. Mathematical representations of the outcomes of these processes\footnote{Various mathematical forms of PDFs for mass functions have been presented by \citet{DeMarchi2010,Myers2011,Chabrier2014,Hoffmann2018,Essex2020}.}can closely resemble one another, for example figure 2 of \citet{Basu2015} shows a close convergence of the MLP to the Chabrier IMF for suitable parameters, as does the turbulent fragmentation model presented by \citet{Chabrier2014}. Hence, a fitting of a function to an observed MF cannot in itself determine the stellar formation mechanism.   

In this paper, we apply the MLP distribution to the investigation of the mass function (MF) of a resolved star cluster. The latter, having largely the same chemical composition and age, are assumed to be simple stellar populations \citep[SSPs; e.g.,][]{Niederhofer15}, making them ideal targets for IMF studies. The IMF of an SSP can be different from the MF because of various effects such as dynamical evolution, mass segregation, and the presence of unresolved binaries. The aim of our study is to determine the underlying shape of the mass distribution using the MLP. We also focus on understanding how the underlying fitting techniques and the range of data being fitted can introduce systematic errors when fitting the MF, and by extension to the IMF. Here, we present a pilot study introducing our method and its application to NGC 1711, a stellar cluster located in the Large Magellanic Cloud (LMC). The LMC is a gas-rich satellite galaxy of the Milky Way located at a distance of $\sim 50$ kpc \citep{Freedman01}. It is well within the virial radius of the Milky Way, and has been interacting with the Small Magellanic Cloud \citep{Putman03}. This interaction has led to a burst of star cluster formation \citep{Harris09,Nidever10}. Overall, the star clusters in the LMC span a wide range in ages ($10^6 - 10^{10}$ yr) and masses ($10\,M_{\odot} - 10^6\,M_{\odot}$) \citep[e.g.,][]{Hunter03}. Stellar clusters in the Magellanic Clouds have been studied extensively due to their proximity and as a means to calibrate stellar evolutionary models, among other science cases. Therefore, a rich dataset of observations of a large number of stellar clusters with the Hubble Space Telescope (HST) exist with similar observational characteristics to those we study here for NGC 1711 \citep[e.g.,][]{Mackey2003}. NGC 1711 is located in the northwest part of the LMC, below its bar. It is a populous young star cluster, with an age of $10^{7.70\pm 0.05}$ yr, a metallicity of $-0.57\pm 0.17$ dex and a reddening $E(B-V)$ of $0.09\pm0.03$ \citep{Dirsch00}. 

Our goal is to introduce a general approach by which to determine whether the MF of a resolved stellar system can be best described by a power law, a lognormal, or a hybrid function that has features of both. We promote the use of the MLP distribution as an efficient means of accomplishing this. We use the case of NGC 1711 to also illustrate how the availability of observational depths resulting in mass ranges that do not probe the lognormal part of the IMF can affect the interpretation of the MF, including the index of the power-law tail. We also seek to understand variations in the MF properties that result from different data analysis techniques. Our goal is not to find evidence that one or the other of accretion termination or turbulent fragmentation is a preferred means for the generation of the MF, but rather to use the MLP to explore the shape of the mass distribution and the possible random and systematic errors when fitting the underlying data set. This paper is organized as follows. In Section \ref{MLPsection} we introduce the MLP function. Section \ref{Observations} describes the observations and data analysis. Section \ref{Results} contains results of the fitting of the mass function using the MLP through nonlinear regression as well as maximum-likelihood estimation (MLE). Section \ref{Summary} contains a summary of results.

\section{The MLP distribution} \label{MLPsection}
% describe the derivation of the MLP in words

The mass of a protostellar condensation depends on many physical parameters and its multiplicative dependence on the latter provides the condensations with an initial lognormal mass function (according to the CLT). The condensation may initially gain mass from its surroundings at a constant accretion rate \citep{Shu1977} that can be modulated by a decline in mean value \citep[e.g.,][]{Foster1993,Basu2005} and an episodic manner \citep[e.g.,][]{Basu2006, Basu2010, VorobyovBasu2015}. However, when massive stars form, their formation time does not appear to be much longer than that of surrounding low-mass stars \citep{Myers1993}. Massive star formation seems to require a period of rapid accretion perhaps channeled by flows from the molecular cloud \citep{Wang2010,Myers2011,Myers2014}.

One can incorporate some of these features using a simple exponential growth formula that initially has slow growth but grows rapidly once past a characteristic growth time scale. Furthermore, if one also assumes an exponential distribution of accretion times, then most condensations may terminate accretion growth before the accretion rate rises and leads to massive star formation. \citep[][see also \citet{Reed2003}]{Basu2004} show that the result of these two assumptions is a PDF for final masses with a lognormal body and a power-law tail, in which the power-law index is the ratio of the growth time of the accretion rate to the termination time scale of the exponential distribution of lifetimes. The PDF that is obtained by integrating over time is a single function having a lognormal body and a power-law tail and is called the MLP.

The MLP function is a three-parameter PDF. If $m$ is the mass of a star, the PDF of the MLP function is given in closed form as 
\begin{eqnarray}
f(m) & = & \frac{\alpha}{2}\exp\left(\alpha\mu_0 +\alpha ^2\sigma_0 ^2 / 2\right)m^{-(1+\alpha)} \nonumber \\ 
& \times & \text{erfc} \left(\frac{1}{\sqrt{2}}\left(\alpha\sigma_0 - \frac{\ln m - \mu _0}{\sigma_0}\right)\right), \, m \in [0,\infty) \label{MLP} 
\end{eqnarray}
\citep{Basu2015}. The three parameters describing the MLP distribution function are $\alpha$, $\mu_{0}$, and $\sigma_{0}$: $\alpha$ represents the power-law tail, while $\mu_{0}$ and $\sigma_{0}$ describe the shape of the lognormal body. While $\alpha$ is the power-law index, which is also characteristic of the pure power-law Pareto distribution, $\mu_{0}$ and $\sigma^{2}_{0}$ do not represent the mean and variance of the distribution as in the lognormal distribution; we provide summary statistics for the MLP below. It is important to note that the MLP behaves as a pure power-law distribution in the limit of $\sigma_{0}$ tending to zero. The functional forms for the lognormal and the Pareto distributions are, respectively,

\beq 
f(m) = \frac{1}{\sqrt{2\pi}\sigma\,m} \exp{\Bigg[-\frac{(\ln\,m - \mu)^2}{2\sigma^2}}\Bigg] \ ,\label{lognormal} 
\eeq
and
\beq
f(m) = Am^{-(1 + \alpha)},
\eeq
where $A$ is a scaling parameter.
\\
\\Some properties of the MLP function are as follows.\\
\\
(i) Mean:
\beq 
E[M] =\frac{\alpha}{\alpha - 1} \exp\left(\frac{\sigma_0^2}{2} + \mu_0 \right) \label{MLPMoments} , \ \alpha > 1; 
\eeq
(ii) Variance: %
\beq 
\label{var}
\text{Var}(M) = \alpha \exp(\sigma_0^2 + 2\mu_0)\left(\frac{e^{\sigma_0^2}}{\alpha-2} -  \frac{\alpha}{(\alpha - 1)^2}\right), \ \alpha > 2;
\eeq 
(iii) Cumulative Distribution Function:
\beq 
F_M(m) = \frac{1}{2} \text{erfc} \left(-\frac{\ln m-\mu_0}{\sqrt{2}\sigma_0}\right)-\frac{1}{2}\exp\left(\alpha\mu_0+\frac{\alpha^2\sigma_0^2}{2}\right) \\ 
m^{-\alpha} \text{erfc} \left(\frac{\alpha\sigma_0}{\sqrt{2}}-\frac{\ln m-\mu_0}{\sqrt{2}\sigma_0}\right). %\label{cumDist} 
\eeq

\section{Observations and Analysis}
\label{Observations}

We derive an MF from HST/Wide Field Planetary Camera 2 (WFPC2) archival observations of NGC 1711, retrieved through the Mikulski Archive for Space Telescopes (MAST)\footnote{\url{https://archive.stsci.edu/hst/search.php.}}. This data set is a part of the HST program GO-5904 \citep{Fischer98}, consisting of pre-reduced data sets with exposure times 2$\times$300 s in F555W\,($\sim V$) and F814W\,($\sim I$), as well as $2 \times10$ s in F555W and $1 \times 10$ s in F814W. 

We perform point-source photometry to the imaging data set with HSTphot \citep{Dolphin00}, a photometry package specifically designed to handle HST/WFPC2 observations. The detailed photometric techniques used here are identical to those used and described in \citet[][]{Lianou13,LianouCole2013,Spetsieri2018}. The final product of the \textit{hstphot} photometry are two science catalogs, one for the short and one for the long exposure data set. These two photometric catalogs are further combined into one, taking into account that some stars may have been detected in either catalog, so as not to include them twice. Magnitudes are provided in both the WFPC2 and the Landolt UBVRI photometric systems and we choose to use the latter for our study. The combined catalog contains 10363 stars, ranging from 15.3 to 26.2 mag in the $V$-band. %The standard errors from the photometry range from 0.05 mag to 0.2 mag for the V and I magnitudes. 

We quantify the incompleteness of the data conducting artificial star tests, using the utilities provided in HSTphot as in \citet[][]{LianouCole2013}. In this way, two simulated fits images, one for each filter $V$ and $I$, are created using as inputs the coordinates and magnitudes of the stars from the science catalog of our photometry. For these two simulated images, we perform point-source photometry in the same way as we did for the observations, in order to derive positions and magnitudes of the simulated stars, i.e. an output simulated photometric catalog. The ratio of the detected stars in the output photometric catalog to those from the input photometric catalog defines the incompleteness factor, while their difference provides the measure of the photometric uncertainties based on the artificial star tests. 

The incompleteness factor characterizing the star counts depends on both the distance of the detected stars from the center of the cluster (usually placed at the center of the camera) and on their magnitude \citep[e.g.,][]{LianouCole2013}. The incompleteness factor will increase toward fainter magnitudes, due to the  detection limit for the fainter stars \citep[e.g.,][]{Harris1990}. Moreover, crowding effects in the central regions of the cluster will hamper the detection of faint stars there. We use our artificial star tests to quantify these effects globally for the star counts in the field of view. For our analysis, we use MS stars with magnitudes characterized by a small incompleteness factor, such that the completeness of our data is greater than 90\% at the same magnitudes. This occurs at 23 mag in the $I$-band, hence we choose stars brighter than this to include them in our analysis. Given our choice to include stars with magnitudes characterized with a very small incompleteness factor (less than 10\%), we do not perform incompleteness corrections to our data, as this is not altering the star counts per magnitude bin when constructing the global LF/ MF. We consider the global incompleteness of our data, as the aim is to characterize the global LF/MF, rather than how the latter varies as a function of the distance from the center of the cluster, i.e., mass segregation effects. As mass segregation involves the dynamical evolution of the massive stars within a dense cluster, its investigation via the radial variation of the LF/MF is out of the scope of the present study, while it is not required for the interpretation of our global findings on the LF/MF.   

\subsection{Luminosity and Mass Function}\label{LF}
%%%%%% FIGURE 1- isochrones  %%%%%%%%%% 
	\begin{figure} 
	\centering
	\includegraphics[width = 0.5\columnwidth]{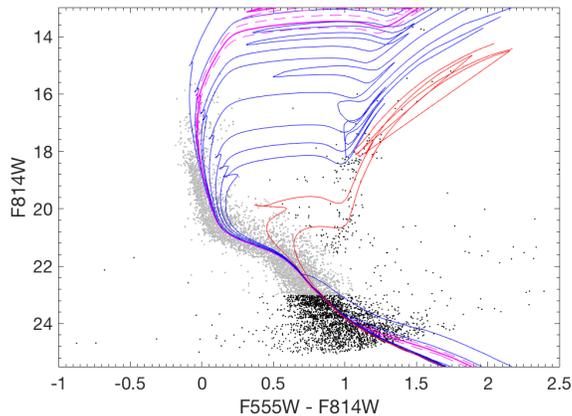} 
	\caption{Color-magnitude diagram, F814W  vs. F555W -- F814W. The PARSEC isochrones are shown with blue lines corresponding to ages of 20, 40, 80, 100, 200, 400, 600, and 800 Myr from top to bottom, and with red lines corresponding to ages of 2 and 4 Gyr. The magenta lines correspond to isochrones of $\log (t/{\rm yr}) = 7.70 \pm 0.05$ using PARSEC stellar tracks. The gray points represent the MS stars with incompleteness factors less than 90\% in the $I$-band.}
        \label{isochrones}
	\end{figure}

The LF is the distribution of stellar absolute magnitudes in a particular wavelength (here probed with the F814W-band) into different absolute magnitude intervals $[M_{i},M_{i} +dM_{i}]$. In order to obtain the LF, we correct the data for field star contamination. 
%For field star contamination, we first plot the spatial distribution of the stars to then select those stars that are located in the outskirts of the stellar cluster probed within the HST field-of-view. We choose the outermost regions of the stellar cluster probed within the HST field-of-view to best describe the field, and for a radially declining stellar distribution for the cluster member stars, we assume that most of the stars in the outer regions belong to the field rather than the cluster. 
To do this, we assign as field stars those stars located in the outermost region of the stellar cluster that is probed with the HST field of view. Given the radially declining number density of cluster member stars, the stars in this outermost region are assumed to belong to the field rather than the cluster.
We then plot the histogram of the apparent magnitudes for the selected field stars and subtract it from the histogram of the apparent magnitudes for stars in the stellar cluster, to get the new counts representing the stellar cluster members \citep{Mateo1988,Sagar1991}. We do not perform corrections for binary stars as these do not affect the slope of the MF significantly \citep{Zeidler2017}. The color magnitude diagram (CMD) for the cluster is shown in Figure~\ref{isochrones} with overlaid isochrones. The overlaid isochrones are derived from theoretical models to obtain the mass-magnitude relationship (MMR). We first divide the $M_{\rm F814W}$ absolute magnitude into bins of optimal size $2n^{2/5}$, where $n$ is the total number of points (i.e., 4177 stars) to obtain the LF \citep{Maschberger2009}. We choose stars on the main sequence to have a one-to-one correspondence between absolute magnitude and mass for a well defined MMR \citep{Sirianni2000}. 

%%%%% FIGURE 1 - CMD  %%%%%%%%%% 

	%\begin{figure}
	%\centering
        %\includegraphics[trim=1cm 0cm 1cm 1cm,width=8.5cm,clip]{figs/1CMD1711.pdf}
        % trim= left bottom top right
	%\caption{Color Magnitude Diagram of NGC 1711. On the horizontal axis is color, i.e. F555W - F814W, and on the vertical axis is the apparent magnitude F814W. The red circles represent the background field stars.}
       %\label{dm_figure1}
      % \end{figure}

%%%%%%%%%%%%%%%%%%%%%%%%%%%%%%%%%%%       
%% 
     % do sirriani and sabbi here instead. % you also do hunter holtzmann or put them under completeness correction

%%%%% FIGURE 2 - LF  %%%%%%%%%% 
	%\begin{figure} 
	%\centering
        %\includegraphics[trim=2.9cm 0.5cm 2.9cm 1cm,width=4.1cm,clip]{figs/LF.pdf}
       % \includegraphics[width = 0.50\textwidth]{figs/2Apparentmagnitude.pdf}
	%\caption{Number of stars distributed as a function of their apparent F814W-band magnitude. The magnitude range is $15.31 < F814W <  22.80$. }
      % \label{dm_figure2}
	%\end{figure}

%%%%% FIGURE 2 - LF log  %%%%%%%%%% 
	\begin{figure} 
	\centering
	\includegraphics[width = 0.50\textwidth]{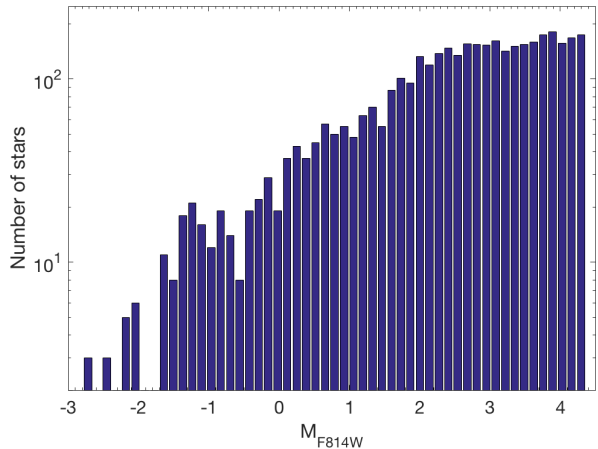}
	\caption{Number of stars distributed as a function of their absolute F814W-band magnitude, $M_{F814W}$.}
         \label{AbsMag}
	\end{figure}
	
%%%%%%%%%%%%%

Figure~\ref{AbsMag} shows the distribution of stars as a function of their absolute F814W-band magnitude, i.e., the LF. The brightest MS star in NGC 1711 has an apparent F814W-band magnitude of 15.31\,mag, or -3.05 absolute magnitude. We make the conversion from apparent to absolute magnitude using the distance modulus of 18.25 mag and an $I$-band foreground extinction of 0.12 mag\footnote{The values for the foreground extinction in the direction of NGC 1711 are taken from NASA/IPAC Extragalactic Database (NED).}. For the $V$-band extinction of 0.21\,mag, the foreground reddening is then 0.09 mag. 

We consider two different sets of theoretical isochrones, i.e., PARSEC \citep{Bressan2012,Chen2015} and MIST \citep[][]{Mesa2013,Choi2016}, to convert from the LF to MF and to investigate possible dependence of the MF on the choice of an MMR.
%	
%%%%% FIGURE 3 - MMR compare  %%%%%%%%%% 	
	%\begin{figure} 
% 	\centering
% 	\includegraphics[width = 0.45\textwidth]{figs/comparisonmass.pdf} 
% 	\caption{Theoretical MMR for log ($t$/yr) = 7.70. The red points represent the MMR obtained using PARSEC isochrones and the blue points represent the one obtained with MIST isochrones.}
%         \label{comparisonmass}
% 	\end{figure}
In Figure~\ref{isochrones}, the PARSEC isochrones are overlaid on the CMD of NGC 1711. We do not obtain the age and metallicity for the stellar cluster anew; rather we adopt the values from the detailed analysis of \cite{Dirsch00} and assign the isochrone corresponding to an age  log ($t$/yr) = 7.70  and the metallicity to be $-0.57\pm0.17$ dex. We use these values for both PARSEC and MIST, in order to derive the relation between mass and absolute magnitude for MS stars. We then interpolate the mass and magnitude values for the given age and metallicity of the cluster to obtain the MMR.

%
%%%%% FIGURE 6 - MF   %%%%%%%%%% 	
	%\begin{figure}
	%\centering
	%\includegraphics[width = 0.50\textwidth]{figs/6MassFunc.pdf}
	%\caption{The probability density $f(m)$ as a function of normalised stellar mass, $m$, in units of $M_{\odot}$. The red line corresponds to PARSEC isochrones, and the blue lines to MIST isochrones.}
	% \label{dm_figure6}
 %\end{figure}
%

%
%%%%% FIGURE 3 - mfm  %%%%%%%%%% 	
	\begin{figure}
	\centering
	\includegraphics[width = 0.50\textwidth]{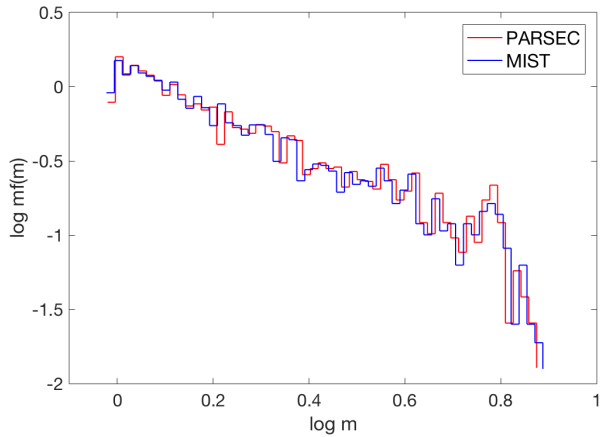}
	\caption{The logarithm of the probability density multiplied by mass, $\log mf(m)$, as a function of the logarithm of the normalized stellar mass, $\log m$, in units of $M_{\odot}$ for the stellar mass. }
	 \label{logMF}
 \end{figure}
The MMR obtained from the two isochrone models have similar behavior, although probing slightly different mass ranges. The masses obtained from the PARSEC isochrones span the range 0.90 $M_{\odot}$ to 7.63 $M_{\odot}$, while those obtained from MIST range from 0.90 $M_{\odot}$ to 7.87 $M_{\odot}$. In figure~\ref{logMF}, we plot $m f(m) = \Delta N'/ \Delta \log m$ i.e. the derived MF, where $\Delta N' = \Delta N/ N_{\rm total}$ and $\Delta \log m = 0.03$, on the vertical axis with $\log m$ on the horizontal axis. We bin the data to the same optimal bin number (55) as used for the LF. The sudden decline in number of stars above $\sim$6\,$M_{\odot}$, seen in the MFs obtained using either of the isochrone models, is due to stellar evolution and/or the stochastic nature of sampling the high-mass stars \citep[][and references therein]{LianouCole2013}. %This is discussed in Section~\ref{Highmass} in more detail. 

%\section{MLP modeling of the Mass Function}
\section{MLP model fitting}\label{Results}

In this section, we use parametric model fitting to model the MF of NGC 1711. Usually the method of linear/nonlinear regression is used for model fitting. Regression is a statistical technique that analyzes the relationship between a dependent variable and several independent variables \citep{babu2012}. The main objective of regression/nonlinear least-squares method is to estimate the unknown parameters of the mathematical equation by minimizing the sum of squares of the residuals \citep{motulsky1987fitting,johnson1992and}. In this paper, we use the Levenberg-Marquardt (LM) method on the normalized MF in order to estimate the best-fit parameter values of the MLP function. 

The LM method is an iterative process evolved from the combination of the Gauss-Newton and the steepest descent method. It uses the advantages of each of the two methods to compute best estimates for the parameters of the given equation. The method of steepest descent is advantageous in initial iterations as it quickly moves along the direction of steepest descent to minimize the sum of squares of the residuals, but it becomes less accurate on later iterations. Unlike the method of steepest descent, the Gauss-Newton method is effective for later iterations but may go in the wrong direction for initial iterations. Hence, the LM method jumps from the steepest descent to the Gauss-Newton method from initial to later iterations \citep{levenberg1944method}. Like the method of steepest descent and the Gauss-Newton method, the LM method is an iterative process and requires an initial estimation of the parameters. From these, it tries to find a better estimate to the parameters by minimizing the sum of the squares of the residuals. To check whether the algorithm gives the best fitting parameters, it is important to understand how good the fit is and how much uncertainty is involved. 

%are the assumptions an an if and only condition? 
The results from nonlinear regression are robust if the underlying assumptions for the least-squares approach are satisfied \citep{zielesny2011curve}. There are two underlying assumptions:  the uncertainties involved in the fitting should be randomly distributed and must follow a normal distribution. 

{\subsection{Nonlinear Regression} 

To do the fitting we use the LM algorithm for the MLP function on the MF obtained from the two theoretical isochrones, i.e., PARSEC and MIST, shown in Figures 5 and 6, respectively.  For fit 1 (using PARSEC MMR) the method converges to $\alpha = 1.67$, $\mu_0 = -0.06$ and $\sigma_0 = 0.07$. For fit 2 (using MIST MMR) we obtain $\alpha = 1.73$, $\mu_0 = -0.07$ and $\sigma_0 = 0.04$. It is important to note that the parameter $\sigma_0$ for the MLP function lies closer to zero for the above best-fit set of values. This implies that the MLP behaves as a pure power law in this limit of $\sigma_0$ tending to 0. 

For fit 1 (PARSEC), we obtain a root mean square error (RMSE) $= 0.17$. We further used the same algorithm to fit a pure power-law distribution, which is already a candidate model for the MF, and found that it has the same RMSE value and the same slope value as well i.e., $\alpha = 1.67$.  Similarly, for fit 2 (MIST), we obtain RMSE $= 0.16$. 
For fit 1, we obtained 95\% confidence bounds for each parameter of the MLP as $(1.49, 1.87)$ for $\alpha = 1.68$, $(-0.15, 0.02)$ for $\mu_0 = -0.06$, and $(-0.09, 0.23)$ for $\sigma_0 = 0.07$. For fit 2, these are $(1.55, 1.90)$ for $\alpha = 1.73$,  $(-0.15,0.01)$ for $\mu_0 = -0.07$, and $(-0.08, 0.16)$ for $\sigma_0 = 0.04$.

%%%%% FIGURE 4 - padovafit   %%%%%%%%%% 	
        \begin{figure}
	\centering
	
	\includegraphics[width={0.50\textwidth}]{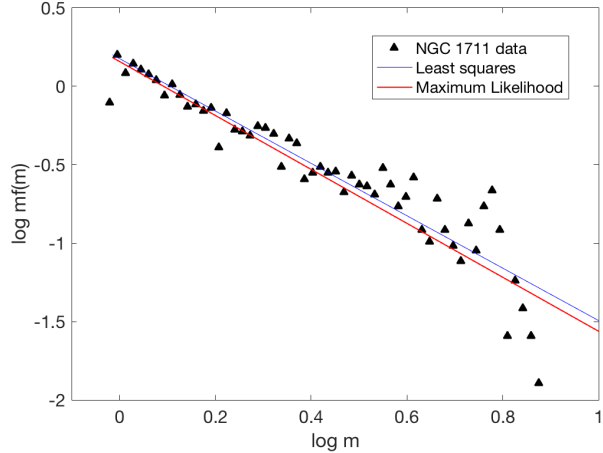}
	
	\caption{\label{Padova} Fits to the MF found using PARSEC isochrones.  The best fit MLP function using least squares is the blue line and has parameters $\alpha = 1.67$, $\mu_0 = -0.06$ and $\sigma_0 = 0.07$. The red line represents the best fit MLP function using maximum-likelihood estimation, for which we find $\alpha = 1.72$. } 
	\end{figure}

%%%%% FIGURE 5 - MIST fit   %%%%%%%%%% 	
	\begin{figure}
	\centering
	
	\includegraphics[width={0.50\textwidth}]{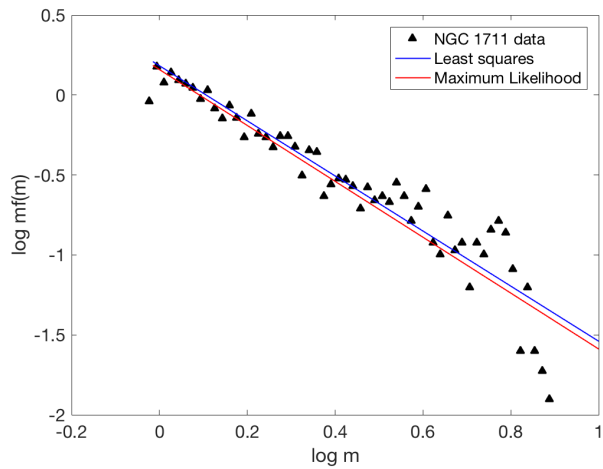}
	
	\caption{\label{MESA} Fits to the MF found using MIST isochrones. The best fit MLP function using least squares is the blue line and has parameters $\alpha = 1.73$, $\mu_0 = -0.07$ and $\sigma_0 = 0.04$.  The red line represents the best fit MLP function using maximum-likelihood estimation, for which we find $\alpha = 1.75$.}  
	\end{figure}
	
%%%%% FIGURE 10 - residuals  %%%%%%%%%% 	
	%\begin{figure}
	%\centering
	%\includegraphics[width = %0.50\textwidth]{figs/10residuals.pdf}
	%\caption{  Residual plot for the best fit MLP function %on the MF obtained using PARSEC isochrones.}  
	%\label{Residual}
	%\end{figure}
	
%%%%% FIGURE 11 - QQ plot   %%%%%%%%%% 	
	 %\begin{figure}
	%\centering
	%\includegraphics[width = %0.50\textwidth]{figs/11QQplot.pdf}
	%\caption{  Quantile-Quantile plot for MF obtained using %PARSEC isochrones. The red line represents the standard %normal.} 
	%\label{QQplot}
	%\end{figure}
%

Since we use a nonlinear regression approach to fit the MLP to the data for NGC 1711, it is important to check whether the underlying assumptions for the least-squares approach are satisfied \citep{zielesny2011curve}. To do so we check whether the uncertainties involved in the fitting are random and normally distributed, i.e. the residuals obtained do not vary systematically above or below zero and follow a Gaussian distribution. After doing a runs test~\citep{motulsky1987fitting} on the residuals we find that the residuals are actually systematically distributed. By performing the Kolmogorov-Smirnov (KS) test~\citep{massey1951, babu2012} we find that the uncertainties are not normally distributed. Violation of the two underlying assumptions affects the certainty of the best-fit parameter values and also the confidence interval for this set of parameters.

\subsection{MLE}\label{ML}
\cite{Maiz2005} showed that deriving the slope for a power-law distribution using a least-squares minimization method with uniform binning of data has significant numerical bias. The correlation between the number of stars in each bin and the weights assigned to each bin causes a bias in the determination of the slope. Hence, there is a need to reinvestigate the values for the model parameters using a method that is independent of binning. To do so, there are methods such as the maximum-likelihood estimation (MLE) and Bayesian parameter estimation \citep{bock2013}. Bayesian parameter estimation treats the unknown model parameters as random variables, thus requiring one to specify a prior PDF to compute a posterior PDF for the model parameters~\citep[e.g.,][]{Dib2014}. The MLE, on the other hand, follows a frequentist approach and treats the unknown parameters as constants instead of random variables, thus not requiring any prior information.

 Since our underlying stellar mass distribution is fixed, it is easier to treat the unknown parameters of the MLP as unknown constants and simply use the method of maximum likelihood to obtain the best-fit parameters. Given a sample of observations $x_1,\,x_2,\,x_3,\,.....,\,x_n$, where the $x_i$'s are independent and identically distributed data points assumed to be taken from a PDF $f(X\,|\Theta)$ of $k$ unknown parameters $\theta_1,\,\theta_2,\,.......,\,\theta_k$, the likelihood function can be defined as
\begin{equation}
L(\Theta\,|x_i) =  f(x_1\,|\Theta) f(x_2\,|\Theta)  ..... f(x_n\,|\Theta) = \prod_{i=1}^{n} f(x_i\,|\Theta) 
\end{equation}
\citep{johnson2002}. Maximizing the likelihood function helps to find the parameter values that are most likely to describe the data set. For simplicity the log of the likelihood function is maximized. One can find the maximum-likelihood estimator for the parameters $\theta_1,\,\theta_2,\,.......,\,\theta_k$ by simultaneously solving 
\begin{equation}
\dfrac{d\,\ln\,L(\Theta\,|x_i)}{d\theta_j}\,=\,0 \, : j = 1,....,k\,.
\end{equation}
For various distributions like the lognormal distribution or the Pareto distribution, functional forms can be found for the maximum-likelihood estimators. For distributions that do not have a functional form for the estimators, global optimization techniques such as simulated annealing~\citep{kirkpatrick1983} or particle swarm~\citep{eberhart1995} can be explored to find global minima for the negative-likelihood function i.e., $ - \ln\,L(\Theta\,|x_i)$, which is same as finding global maxima for the likelihood function. Using simulated annealing, we find maximum-likelihood estimators for the parameters of the MLP to be $\alpha = 1.72 \pm 0.05$, $\mu_0 = -0.10 \pm 0.01$ and $\sigma_0 = 0.01 \pm 0.01$ for fit 1. These lie within the predicted 95\% confidence bounds of the parameter values obtained using regression. For fit 2 we get $\alpha = 1.75 \pm 0.05 $, $\mu_0 = -0.11 \pm 0.01$ and $\sigma_0 = 0.01 \pm 0.01$, which also lie in the predicted 95\% confidence bounds of the parameter values obtained using regression. Also, the maximum-likelihood estimates of fit 1 lie in the 95\% confidence interval of the maximum-likelihood estimates of fit 2 and vice versa. Thus we can conclude that the derived slope, i.e., $\alpha$, is independent of the underlying MMR. 
 
It is interesting to note that our best-fit power-law index falls in between two previous determinations for NGC 1711. \citet{Mateo1988} used $BV$ CCD photometry using the Cerro-Tololo Inter-American Observatory 4 m and 0.9 m telescopes to find a slope $\alpha = 2.4 \pm 0.4$, while \citet{Sagar1991} used $BV$ CCD photometry using the 1.54 m Danish telescope at the European Southern Observatory and found slope values $\alpha = 1.3 \pm 0.2$ or $1.9 \pm 0.3$ depending on which of two theoretical isochrones were used. The difference between the results of \citet{Mateo1988} and \citet{Sagar1991} can be partially attributed to the use of different theoretical isochrones, as discussed by \citet{Sagar1991}. Our results are largely independent of the two isochrones that we have utilized. However, the wide range of values of $\alpha$ obtained in these studies point to systematic uncertainties. While some of these are likely part of the conversion process from an LF to an MF (e.g., the particular isochrone that is used), another part may be due to the fitting process of the MF itself. Both \citet{Mateo1988} and \citet{Sagar1991} used a least-squares fit of a straight line to data that was logarithmically binned. In the next section, we explore the effect of the range of masses that are modeled and of fitting with the hybrid MLP function. The latter  process is more nuanced than fitting a straight line to the mass range.

 %\subsection{High mass end}\label{Highmass}

%SB comment
%Finally, we note that the masses above 6 $M_{\odot}$ in  Figure~\ref{Padova} and Figure~\ref{MESA} fall significantly below the best fit power-law profiles. On further investigation  we find that the confidence intervals for $\alpha = 1.78 \pm 0.05$ using PARSEC MMR and $\alpha = 1.81 \pm 0.05$ using MIST MMR obtained for the distribution of masses below 6 $M_{\odot}$ overlap with the confidence intervals for $\alpha$ obtained for the entire distribution. This shows that $\alpha$ is independent of the masses/data points lying above 6 $M_{\odot}$.   

%The deviation above 6 $M_{\odot}$ may be due to the evolution of stars off the main sequence. For stars above 6.7 $M_{\odot}$ the main-sequence lifetime of the star is less than the estimated 50 Myr age of the cluster.
%so it is expected that these stars move away from the main-sequence branch. 
%Even though all stars above 6.7 $M_{\odot}$ should have evolved off the main sequence, we see a decline in the MF but not a sharp drop to no objects at all. This can be due to the presence of blue stragglers i.e. stars of lower masses that acquire mass through binary mass transfer \citep{Gosnell2015, anna2016} or mergers, or it could represent an age spread of star formation. We do not investigate these phenomena in this paper.

\subsection{MLP as a hybrid}\label{MLPchabrier}

Another significant purpose of using the MLP distribution function is to check whether it can work as a hybrid and model both lognormal as well as power-law behavior as a single function. We also check whether the addition of a lognormal body has any effect on the exponent of the power-law tail. For simplicity, we use only the PARSEC isochrones. Since the observational depth of our data allows us to probe stellar masses only as low as 0.90 $M_{\odot}$, we combine the NGC 1711 cluster data with an artificially generated data sample from the lognormal functional form assuming $\mu_0 = \log_{10}0.2$ and $\sigma_0 = 0.55$, taken from the most commonly used \citet{Chabrier2005} representation of the IMF. We generate 16631 data points in the mass range 0.06 $M_{\odot}$ to 0.90 $M_{\odot}$ from the Chabrier functional form. The total number of data points needed is found by dividing the total number of NGC 1711 stars in the range 0.90 $M_{\odot}$ to 1 $M_{\odot}$ by the total probability from the Chabrier functional form over that range. We combine these synthetic data points with the cluster data from 0.90 $M_{\odot}$ to 7.63 $M_{\odot}$ in order to obtain a complete data sample that includes low-mass to high-mass stars. Then we again use MLE on the combined data sample and obtain the parameter estimates anew; see Figure 6. This yields parameter values $\alpha = 2.04 \pm 0.07$, $\mu_0 = -1.10 \pm 0.01$, and $\sigma_0 = 0.55 \pm 0.01$. The artificially generated data points result in a greater value for $\sigma_0$, which is representative of an addition of a lognormal body to the power-law distribution of NGC 1711 data points. Using the MLP properties \citep{Basu2015}, we also find a shift in the mean of the distribution from 2.34 $M_{\odot}$ to 0.76 $M_{\odot}$. The addition of the lognormal body alters the mean of the distribution, which now lies in the low-mass end of the stellar regime.  Our main aim of joining the NGC 1711 data points with Chabrier data points is to check whether the exponent of the power-law tail is affected by an unseen lognormal body. From our fitting results we obtain the exponent for the combined data to be $\alpha = 2.04 \pm 0.07$, which is steeper than the slope $\alpha = 1.72 \pm 0.05$ found for the NGC 1711 data points alone. This shows that the presence or absence of data from a lognormal body can systematically affect the measured slope of the power-law tail of the distribution. 
%Thus, the addition of the lognormal body does not have a significant effect on the exponent of the power-law tail of the distribution, but it alters the mean of the distribution. 

%
 \begin{figure}
	\centering
	
	\includegraphics[width={0.50\textwidth}]{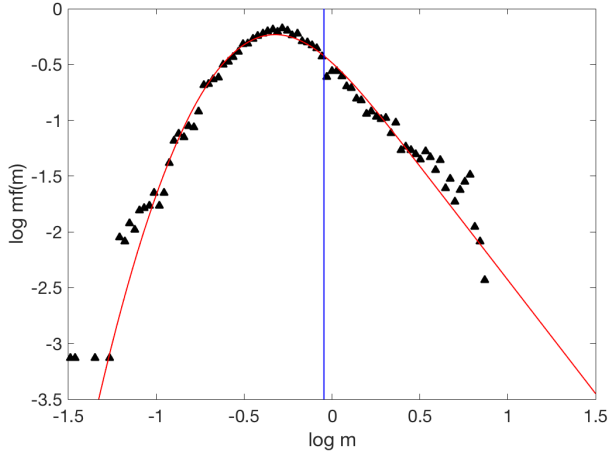}
	
	\caption{MLP fit to the entire stellar mass domain including 
the artificially generated data points from the Chabrier function. 
	The best fit parameter values are $\alpha = 2.04 \pm 0.07$, $\mu_0 = -1.10 \pm 0.01$ and $\sigma_0 = 0.55 \pm 0.01$. The blue line represents the demarcation of the artificially generated data points from the NGC 1711 data points. }
	\end{figure}

%%%% you can include PDMF to IMF and Comparison in the summary itself!!

\section{Summary}\label{Summary}

We have derived the MF for the young and populous LMC stellar cluster NGC 1711 and investigated how the power-law index of the intermediate to high-mass regime can vary when using different analysis techniques. By employing two different sets of theoretical isochrones, we find that the mass range for the MF does not depend on the underlying theoretical MMR, with mass ranges between 0.90 $M_{\odot}$ and 7.63 $M_{\odot}$ with PARSEC, and mass ranges between 0.90 $M_{\odot}$ and 7.87 $M_{\odot}$ using MIST isochrones.}. 

We demonstrated how the MLP function can be employed to test whether the MFs show lognormal, power-law or hybrid behavior.
We found that the MFs follow a pure power-law behavior over the measured mass range and that the MLP function gave the best parameter values for the asymptotic slope $\alpha \equiv d\ln N/d\ln m$ as follows: (i) PARSEC, $\alpha = 1.72 \pm 0.05$;  (ii) MIST, $\alpha =1.75 \pm 0.05$.  In the limit that the parameter $\sigma_0$ of the MLP function tends to zero, the MFs tend to a pure power-law behavior. Indeed, we obtained a very small value $\sigma_0 = 0.02$ for the MFs using either the PARSEC MMR or the MIST MMR. We also found that the least-squares method gives inaccurate results, while the MLE is more robust because it is independent of binning. Hence, we rely on the latter to retrieve the best-fit results to the MLP function. 

The depth of the data available for NGC 1711 have allowed us to probe stellar masses down to 0.90 $M_{\odot}$, hence the lognormal behavior, if any, is not uncovered down to this mass limit. Therefore, we investigated whether the MLP function can successfully model hybrid i.e., lognormal as well as power-law behavior, by adding synthetic data. We generated artificial data points from the \citet{Chabrier2005} lognormal function and combined them with the star counts for NGC 1711 to get a complete data set ranging down to low masses, independent of the available depth of the dataset. Fitting the MLP function using MLE to the MF, using the PARSEC MMR, yielded $\alpha = 2.04 \pm 0.07$, which is steeper than the $\alpha = 1.72 \pm 0.05$ found from fitting the cluster data alone. The absence or presence of a lognormal body in the data can systematically affect the measured power-law tail, and
provide caution to attempts to extend the IMF to extremely high-mass stars based on an index $\alpha$ that is determined from only the intermediate and high-mass regime.

We conclude that the MLP can be used to model hybrid behavior as a single function instead of using different functions with joining conditions. In the case of data that only covers a power-law portion of the mass function, the MLP will naturally converge to a pure power-law fit with a near-zero value of $\sigma_0$. Most importantly, the asymptotic slope $\alpha$ may differ depending on whether the low-mass objects are included in the data set, hence depending on the depth of the observations. We found the value of $\alpha$ is independent of two different isochrone models, but is dependent on the method of fitting. The least-squares fitting technique is still the norm in many MF studies, but we have shown that the underlying assumptions can sometimes be invalid. A comparison of MLE with other techniques such as Bayesian parameter estimation can determine the best method to use in the case of IMF studies. However, the MLE proves to be a better candidate method than least squares, given its independence from data binning.

%%% comparison study

%We use a non-linear least squares regression approach to investigate whether the Modified Lognormal Probability Distribution Function can be used to adequately describe the initial mass distribution of a young and populous simple stellar population (SSP)  NGC 1711. We use the Luminosity Function of the main sequence stars until 23 F814W band-magnitude to avoid considering stars deeper into the main sequence involving significant uncertainties.

\section*{Acknowledgements}
We thank Sayantan Auddy for his valuable inputs. Support for the work of D.M., S.L., and S.B. was provided by the Natural Sciences and Engineering Research Council of Canada. D.M. also thanks Eric Feigelson and Jogesh Babu for conducting the Summer School in Statistics for Astronomers that significantly helped in her research. This research made use of several facilities and open source software: NASA/IPAC Extragalactic Database (NED), operated by the Jet Propulsion Laboratory, California Institute of Technology, under contract with the National Aeronautics and Space Administration; Aladin; NASA's Astrophysics Data System Bibliographic Services; SAOImage DS9, developed by Smithsonian Astrophysical Observatory. %This work made extensive use of the free software GNU Octave and the authors are grateful to the Octave development community for their support.  
%




\end{document}